\documentclass[aps,pre,showpacs,twocolumn,superscriptaddress]{revtex4-1}
\usepackage{times}
\usepackage{graphicx}

\begin{document}

\title{Microwave fidelity studies by varying antenna coupling}

\author{B.~K\"{o}ber}
\affiliation{Fachbereich Physik der Philipps-Universit\"at Marburg, D-35032
Marburg, Germany}

\author{U.~Kuhl}
\affiliation{Fachbereich Physik der Philipps-Universit\"at Marburg, D-35032
Marburg, Germany}

\author{H.-J.~St\"{o}ckmann}
\affiliation{Fachbereich Physik der Philipps-Universit\"at Marburg, D-35032
Marburg, Germany}

\author{T.~Gorin}
\affiliation{Departamento de F\'{i}sica, Universidad de Guadalajara,
Guadalajara C.P. 44840, Jalisco, M\'{e}xico}

\author{D.~V.~Savin}
\affiliation{Department of Mathematical Sciences, Brunel University, Uxbridge UB8 3PH, United Kingdom}

\author{T.~H.~Seligman}
\affiliation{Instituto de Ciencias F\'{i}sicas, Universidad Nacional Aut\'{o}noma de M\'{e}xico, C.P. 62251, Cuernavaca, M\'{e}xico}
\affiliation{Centro Internacional de Ciencias, C.P. 62251, Cuernavaca, M\'{e}xico}

\date{\today
}

\begin{abstract}
The fidelity decay in a microwave billiard is considered, where the coupling to an attached antenna is varied. The resulting quantity, coupling fidelity, is experimentally studied for three different terminators of the varied antenna: a hard wall reflection, an open wall reflection, and a 50$\,\Omega$ load, corresponding to a totally open channel. The model description in terms of an effective Hamiltonian with a complex coupling constant is given. Quantitative agreement is found with the theory obtained from a modified VWZ approach [Verbaarschot~et~al, Phys.~Rep.~\textbf{129},~367~(1985)].
\end{abstract}


\maketitle

\section{Introduction}
\label{sec:Intro}

Fidelity is a standard benchmark in quantum information, and plays a relevant role in discussions on quantum chaos \cite{gor06c}. The corresponding fidelity amplitude can be interpreted as the overlap between two wave functions obtained from the propagation of the same initial state with two different time evolutions or, alternatively, as the overlap of the initial state with itself after being propagated forward in time with one evolution and backward in time with the other. In the latter case one often speaks of Loschmidt echo. Fidelity contains information on both eigenfunctions and spectra of the original and perturbed systems in a non-trivial way. One can show, however, that there exists a profound relation \cite{koh08} between fidelity decay and purely spectral universal parametric correlations \cite{sim93b,tan95,ale98,mar03} in chaotic and disordered systems. The connection holds in quite general settings \cite{smo08}. Efforts to measure fidelity are therefore very important.

There was an early proposal (without the name fidelity) in quantum optics \cite{gar97}. Along these lines the perturbation of a kicked rotor was discussed in great detail \cite{hau05}. A realization of that idea is not available today, although an experiment of this type, but with a more complicated process, was conducted \cite{and03}.

Experiments with microwave cavities or elastic bodies seem to provide good options to study the decay of fidelity \cite{sch05b}, but a difficulty arises. Fidelity implies an integration over the entire space. In two-dimensional microwave billiards the antenna always represents a perturbation, and thus moving the antenna defeats the purpose of a fidelity measurement, as the wave-function taken at any point is that of a slightly different system. In contrast to wave function measurements, in fidelity experiments we are precisely interested in such differences, and thus wave functions measured with moveable antennas \cite{ste92,ste95,kuh07b} or a moveable perturbation body \cite{sri91,bog06,lau07} are not appropriate. In elastic experiments on solid blocks \cite{lob03b,gor06b,lob08} or three-dimensional (3D) microwave billiards the wave function inside the volume seems to be inaccessible anyway \cite{doer98b,alt97a}. This leads to the development of the concept of scattering fidelity \cite{sch05b} which tests the sensitivity of $S$-matrix elements to perturbations. This is also of intrinsic interest since the scattering matrix may be considered  as the basic building  block at least in the case of quantum theory \cite{str00,leh55}.

In former studies the scattering fidelity has been investigated in chaotic microwave billiards by considering a perturbation of the billiard interior. It can be shown that in such a case the random character of wave functions causes the scattering fidelity to represent the usual fidelity, provided that appropriate averaging is taken \cite{sch05b,hoeh08a}. Scars and parabolic manifolds will obviously change that correspondence, but their effect can be avoided in experiment. Specifically, two different types of interior perturbations were experimentally studied. In the first set of experiments a billiard wall was shifted, realizing the so-called global perturbation \cite{sch05b,sch05d}, meaning that there is a total rearrangement of both spectrum and eigenfunctions already for moderate perturbation strengths. Good agreement with prediction from random matrix theory (RMT), expecting Gaussian or exponential decay depending on perturbation strength, was found. In the second experiment a small scatterer was shifted inside the billiard, the wave function being influenced only locally \cite{hoeh08a}. Using the random plane wave conjecture, an algebraic decay was predicted and confirmed experimentally.

Actually, any measurement opens the system. Coupling to the continuum changes drastically the system properties by converting discrete energy levels into unstable resonance states. The latter reveal rich dynamics when the coupling strength to the scattering channels is varied \cite{sok92}, see also \cite{per00} for relevant microwave studies. Since the time evolution operator is subunitary in this case, there appears the leakage of the norm inside the scattering system \cite{sav97}. This decay is fully controlled by the degree of system openness and may also be considered as a remote analog of fidelity decay for open systems. In the framework of the scattering fidelity coupling to the continuum can be taken into account naturally.

It seems therefore attractive to study the sensitivity of $S$-matrix elements to perturbations in the coupling between the scattering system and decay channels. This will be the central purpose of the present paper. Experimentally, we realize the system by a flat microwave billiard with two attached antennas and measure the reflection in one antenna while modifying the coupling in another, see Sec.~\ref{sec:exper} for details on the experimental setup. Section~\ref{sec:theory} presents a theoretical consideration based on RMT and the effective Hamiltonian approach. In Sec.~\ref{sec:results} we discuss in detail the experimental results and compare them with the theory. Our main findings are then summarized in the concluding Sec.~\ref{sec:conclusions}.

\section{Experiment}
\label{sec:exper}

The basic principles of billiard experiments with microwave cavities as a paradigm of quantum chaos research are described in detail in \cite{stoe99}. Therefore, we concentrate on the aspects of relevance to the present study. Reflection and transmission measurements have been performed in a flat resonator, with top and bottom plate parallel to each other. The cavity can be considered as two-dimensional for frequencies $\nu\, <\,\nu_{\rm max} = c/(2h)$, where $h=\rm 8\,mm$ is the height of the resonator.

\begin{figure}[t]
\includegraphics[width=.9\columnwidth]{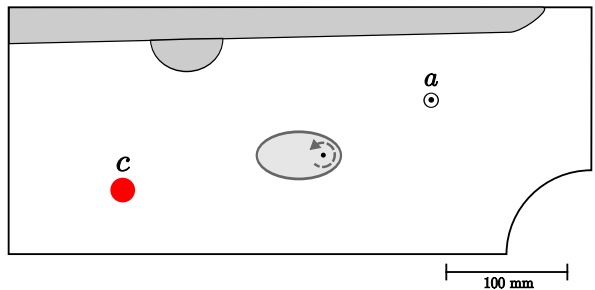}
\caption{\label{fig:01}
Geometry of the chaotic Sinai billiard, length $l=\rm 472\,mm$, width $w=\rm 200\,mm$ and a quarter-circle of radius $r =\rm 70\, mm$ where an antenna with different terminations may be introduced at position $c$. $a$ denotes the measuring antenna. The additional elements were inserted to reduce the influence of bouncing balls.}
\end{figure}

The setup, as illustrated on Fig.~\ref{fig:01}, is based on a quarter Sinai shaped billiard. Additional elements were inserted into the billiard to reduce the influence of bouncing-ball resonances. The classical dynamics for the chosen geometry of the billiard is dominantly chaotic. At position $a$ one antenna is fixed and connected to an Agilent 8720ES vector network analyzer (VNA), which was used for measurements in a frequency range from $\rm 2$ to $\rm 18\,GHz$ with a resolution of $\rm 0.1\,MHz$. We measured the reflection $S$-matrix element $S_{aa}$ first for the unperturbed system, which corresponds to the situation, where no additional antenna is inserted at position $c$. Then we perturbed the system by inserting another antenna at position $c$ which was terminated consecutively in three different ways:
\vspace*{-4ex}

\begin{enumerate}
\renewcommand{\labelenumi}{(\alph{enumi})}
\item connection to the VNA (total absorption),\\[-4ex]
\item standard open (open end reflection),\\[-4ex]
\item standard short (hard wall reflection),\\[-4ex]
\end{enumerate}
and again measured the corresponding reflection at antenna $a$ for each case. The connection of antenna $c$ to the VNA corresponds to a termination of antenna $c$ with a $50\,\Omega$ load. The terminators for the cases (b) and (c) have been taken from the standard calibration kit (Agilent 85052C Precision Calibration Kit) being part of our microwave equipment. For case (a) the reflection amplitude $S_{cc}$ was  also measured. From this measurement the coupling strength of antenna $c$ can be obtained, see Eq.~(\ref{eq:Tc}) below. For all four cases we measured 18 different realizations by rotating an ellipse (see Fig.~\ref{fig:01}) to perform ensemble averages.

An alternative to the coupling of an antenna with variable end is an open wave guide whose coupling to the billiard can be varied by a variable slit. It showed up that, contrary to intuition, for this setup the main effect of the variation of the slit does not correspond to a change of the coupling to the outside, but to a distortion of the wave functions in the billiard, thus corresponding more to the case of a local scattering fidelity \cite{hoeh08a}. This system is discussed in Appendix \ref{app:Exp}.

\section{Theory}
\label{sec:theory}

\subsection{Generalized VWZ approach to fidelity}
\label{subsec:VWZ}

The general case of $M$ scattering channels connected to $N$ levels of the closed cavity can be described in terms of the following effective non-Hermitian Hamiltonian
\begin{equation}\label{eq:Heff}
 H_{\mathrm{eff}} = H - i\sum_{k=1}^{M}\lambda_k V_kV_k^{\dag}\,.
\end{equation}
Here, the internal Hamiltonian $H$ of the closed system is represented by a Hermitian $N\times N$ matrix, whereas $V_k$ are $M$ vectors of length $N$ containing the information on the coupling of the levels to the continuum. The $V_k$ are assumed to be normalized to one, $V_k^{\dag}V_k=1$, and $\lambda_k$ is the coupling constant of channel $k$.

Such an approach was initially developed in nuclear physics \cite{mah69,ver85a,sok89} and since then has been successfully applied to study various aspects of open systems, including wave billiards \cite{stoe99,fyo97b,dit00,fyo05a}. Usually, the phenomenological coupling constants $\lambda_k$ are considered as real numbers which enter the final expressions via the so-called transmission coefficients. However, in the present case of the antenna variation one has to consider the coupling to the variable antenna, $\lambda_{c}$, as a complex number, see discussion in Sec.~\ref{subsec:heff} below. For the sake of generality, we will treat all $\lambda_k$ as complex numbers with the only constraint on their real parts $\mathrm{Re}(\lambda_k)\geq0$, due to the causality condition on the $S$-matrix. We note that quite a similar problem of nonzero $\mathrm{Im}(\lambda_k)$ arises in shell-model calculations due to the principle value term of the self-energy operator, cf.~\cite{mah69} and \cite{ver85a}. This requires proper modification of the theory which we briefly outline below.

According to the general scattering formalism \cite{mah69,ver85a}, the resonance part of the $S$-matrix at the scattering energy $E$ can be expressed in terms of $H_{\mathrm{eff}}$ as follows:
\begin{equation}\label{eq:s_cc1}
 S_{ab}(E) = \delta_{ab} - 2i\sqrt{\mathrm{Re}(\lambda_a)\mathrm{Re}(\lambda_{b})}\, V_a^{\dag}\frac{1}{E-H_{\mathrm{eff}}}V_{b}\,.
\end{equation}
Being interested in a reflection amplitude in channel $a$, it is possible, following \cite{fyo05a}, to obtain another representation for an arbitrary diagonal element $S_{aa}$. To this end, it is convenient first to single out the contribution to $H_{\mathrm{eff}}$ due to channel $a$ by writing $H_{\mathrm{eff}}=H_{\mathrm{eff}}^{a}-i\lambda_a V_a V_a^{\dag}$, and then treat $V_aV_a^{\dag}$ as a rank 1 perturbation to the term $H_{\mathrm{eff}}^{a}=H-i\sum_{k\neq a}\lambda_k V_k V_k^{\dag}$. The upper index ``$a$'' for $H_{\mathrm{eff}}^{a}$ denotes that the contributions of all channels save the given one, $a$, are included. In the next step we expand $(E-H_{\mathrm{eff}})^{-1}$ into a power series with respect to $(E-H_{\mathrm{eff}}^a)^{-1}V_aV_a^{\dag}$ and, after summing up the resulting geometric series, obtain the following general relationship (Dyson's equation) for the corresponding resolvents \cite{sok89}:
\begin{eqnarray}\label{eq:dyson}
 \frac{1}{E-H_{\mathrm{eff}}} &=& \frac{1}{E-H_{\mathrm{eff}}^a} -i\lambda_a\frac{1}{E-H_{\mathrm{eff}}^a}V_a \nonumber \\
 &&\times
 \frac{1}{1+i\lambda_a V_a^{\dag}\displaystyle\frac{1}{E-H_{\mathrm{eff}}^a}V_a}V_a^{\dag}
 \frac{1}{E-H_{\mathrm{eff}}^a}\,.\quad
\end{eqnarray}
This identity, being substituted in Eq.~(\ref{eq:s_cc1}), yields the following expression for the reflection amplitude $S_{aa}$:
\begin{equation}\label{eq:s_cc2}
 S_{aa}(E) = \frac{1 - i\lambda_a^* V_a^{\dag}\displaystyle\frac{1}{E-H_{\mathrm{eff}}^a}V_a}{1 + i\lambda_a V_a^{\dag}\displaystyle\frac{1}{E-H_{\mathrm{eff}}^a}V_a}\,.
\end{equation}

Representation (\ref{eq:s_cc2}) is very convenient to perform statistical averaging. Adopting RMT approach to model intrinsic chaos, we take $H$ from the Gaussian orthogonal ensemble (GOE) of $N\times N$ random real symmetric matrices which is the appropriate choice for the systems with preserved time-reversal symmetry \cite{stoe99,meh91}. The quantities $V_k$ are considered as fixed real $N$-dimensional vectors of unit length. They are also supposed to be mutually orthogonal that ensures the absence of the direct ``fast'' processes (which could be due to a nondiagonal part of the average $S$-matrix) \cite{ver85a}. In the limit $N\to\infty$, for finite $M$, the leading term for the average value of the resolvent is well known to be $\langle[(E-H_{\mathrm{eff}}^a)^{-1}]_{nm}\rangle=[(E/2)-i\sqrt{1-E^2/4}]\delta_{nm}$, where the imaginary part accounts for the famous Wigner's semicircle law. This implies the following result (valid up to the terms of the order of $M/N$) for the average $S$-matrix \cite{ver85a,sok89},
\begin{equation}\label{eq:s_aver}
 \langle S_{ab} \rangle = \frac{1-\lambda_a^*}{1+\lambda_a}\delta_{ab}
\end{equation}
Here, we set $E=0$ as usual. As a result, the transmission coefficient takes the following form:
\begin{equation}\label{eq:Tc}
 T_a \equiv 1- |\langle S_{aa}\rangle|^2 = \frac{4\,\mathrm{Re}(\lambda_a)}{|1+\lambda_a|^2}\,,\quad a=1,\ldots,M,
\end{equation}
which is in agreement with the result of Ref.~\cite{ver85a} obtained by a supersymmetry calculation. It is worth noting that in the case of real $\lambda$ one gets $T=\frac{4\lambda}{(1+\lambda)^2} \in [0,1]$. In the case of purely imaginary $\lambda$ corresponding to perfect reflection, the channel is closed, $T=0$.

\subsection{Coupling fidelity}

We now proceed with the discussion of the scattering fidelity. Its amplitude is defined in terms of the $S$-matrix elements Fourier transformed into the time domain as follows \cite{sch05b}:
\begin{equation}\label{eq:f_ab}
 f_{ab}(t) = \frac{\langle \hat{S}_{ab}(t)\hat{S}_{ab}^{\prime *}(t)\rangle}{ \sqrt{ \langle\hat{S}_{ab}(t)\hat{S}_{ab}^{*}(t)\rangle\langle \hat{S}_{ab}^{\prime}(t)\hat{S}_{ab}^{\prime *}(t)\rangle }}\,.
\end{equation}
The prime indicates a change of the effective Hamiltonian of the original system after a small perturbation for the backward time evolution. Definition (\ref{eq:f_ab}) guarantees that $f_{ab}(0)=1$. Furthermore, an overall decay of the correlation functions due to absorption drops out, provided the decay is the same for the parametric cross-correlation functions in the nominator and the autocorrelation functions in the denominator \cite{sch05b}. The scattering fidelity itself is
\begin{equation}\label{eq:F_ab}
 F_{ab}(t) = |f_{ab}(t)|^2\,.
\end{equation}
Note that in contrast to the original definition of the scattering fidelity, we allow for a change in the channel vectors as well.

As it is explained above, the original idea is to change only the complex coupling strength $\lambda_c$ to one channel $c$, while the measuring is done on one or two different channels $a,b\neq c$. We denote the resulting scattering fidelity by \emph{coupling fidelity}. We present below an exact RMT prediction for this quantity.

The starting point is to apply the convolution theorem for Fourier transforms to Eq.~(\ref{eq:f_ab}) and relate it to the parametric cross-correlation function $\hat{C}[S_{ab},S_{ab}^{\prime *}](t)$ of the $S$-matrix elements in the time domain \cite{note1},
\begin{equation}\label{eq:ss_ft}
 \langle \hat{S}_{ab}(t)\hat{S}_{ab}^{\prime *}(t)\rangle = \hat{C}[S_{ab},S_{ab}^{\prime *}](t)\,.
\end{equation}
We denote the coupling constant for the variable antenna $c$ in forward and backward time evolution by $\lambda$ and $\lambda^\prime$, respectively. (We omit the lower index ``c'' henceforth.) In the case of unchanged coupling, $\lambda=\lambda^\prime$, the autocorrelation function $\hat{C}[S_{ab},S_{ab}^{*}](t)$ is real and its exact expression is obtained from Verbaarschot-Weidenm\"{u}ller-Zirnbauer (VWZ) integral \cite{ver85a} and is given by 
\begin{equation}\label{eq:VWZ}
 \hat C[S_{ab},S_{ab}^*](t) = \delta_{ab} T_a^2(1-T_a) J_a(t) + (1+\delta_{ab}) T_a T_b P_{ab}(t).
\end{equation}
It is convenient to use the parametrization of Ref.~\cite{gor02a} to write down the explicit expressions for the functions $J_a(t)$ and $P_{ab}(t)$, as

\begin{equation}\label{eq:J}
 J_a(t) = 4\mathcal{I} \left[\left( \frac{r+T_a x}{1+2T_a r +T_a^2\, x} +
  \frac{t-r}{1- T_a(t-r)}\right)^2\right]
\end{equation}
and
\begin{eqnarray}\label{eq:P}
 P_{ab}(t) &=& 2\mathcal{I}\left[
  \frac{T_a T_bx^2 + d_{ab}(r)x+ (2r+1)r}{(1+2T_a r + T_a^2 x)(1+2T_b r +T_b^2 x)} \right.\nonumber \\
   && \left. + \frac{(t-r)(r+1-t)}{[1-T_a(t-r)][1-T_b(t-r)]} \right]\,,
\end{eqnarray}
where
$$x \equiv \frac{2r+1}{2u+1}u^2, \quad d_{ab}(r)\equiv T_aT_b+(T_a+T_b)(r+1)-1$$
and the shorthand $\mathcal{I}$ stands for the integral,
\begin{eqnarray}\label{eq:I}
 \mathcal{I}[\cdots] &=& \int_{\max(0,t-1)}^t \!\!{\textrm d} r\int_0^r{\textrm d} u
  \frac{(t-r)(r+1-t)}{(2u+1)(t^2-r^2+x)^2} \nonumber \\
  & & \times \prod_{k=1}^M\frac{1-T_k (t-r)}{\sqrt{1+2T_k r + T_k^2 x}} [\cdots]\,.
\end{eqnarray}
Here and below, $t$ denotes the dimensionless time measured in units of the Heisenberg time $t_H=2\pi\hbar/\Delta$, where $\Delta$ is the mean level spacing.

The calculation of the correlator [Eq.~(\ref{eq:ss_ft})] in the case of $\lambda \neq \lambda^\prime$ proceeds along the same lines as in \cite{ver85a}, see Appendix \ref{app:theo}. The result turns out to be formally given by the same VWZ expression (\ref{eq:VWZ}), where the transmission coefficient $T_c$ in the varied channel $c$ has to be substituted by
\begin{equation}\label{eq:Teff}
 T_c^{\mathrm{eff}} = \frac{2\left(\lambda+\lambda^{\prime *}\right) }{ \left(1+\lambda\right)\left(1+\lambda^{\prime *}\right)}\,,
\end{equation}
while performing the integration [Eq.~(\ref{eq:I})]. The quantity $T_c^{\mathrm{eff}}$ may be considered as an effective transmission coefficient due to a parametric variation of the coupling strength in the channel $c$. Only if $\lambda=\lambda^\prime$, $T_c^{\mathrm{eff}}$ becomes equal to the conventional transmission coefficient [Eq.~(\ref{eq:Tc})]. In contrast to Eq.~(\ref{eq:Tc}), $T_c^{\mathrm{eff}}$  is generally complex and also $T_c^{\mathrm{eff}}\neq1-\langle S_{cc}\rangle \langle S_{cc}^{\prime*}\rangle$. We note, however, that Eq.~(\ref{eq:Teff}) can be cast in the following form
\begin{equation}\label{eq:Teff2}
 T_c^{\mathrm{eff}} = 1 - S_{\mathrm{eff}}' S_{\mathrm{eff}}^*\,,
\end{equation}
where
\begin{equation}\label{eq:Teff3}
 S_{\mathrm{eff}}' = \frac{ 1-\lambda^{\prime *} }{ 1+\lambda }\,, \qquad
 S_{\mathrm{eff}}^* = \frac{ 1-\lambda }{ 1+\lambda^{\prime *} }\,.
\end{equation}
These quantities might be interpreted as the (average) parametric $S$-matrix amplitudes in the varied channel for the forward and backward time evolution, respectively.

The subsequent evaluation of coupling fidelity cannot be done analytically and will be performed numerically.

\subsection{Effective Hamiltonian description}
\label{subsec:heff}

The experimental situation shall now be mapped onto the theory derived in the preceding subsection. Though the calculation is straightforward, and similar approaches can be found elsewhere \cite{stoe02c}, it is repeated here for the reader's convenience. Let us start with the expression of the scattering matrix in terms of Wigner's reaction matrix:
\begin{equation}\label{eq:s02}
    S=\frac{1-i W^\dag GW}{1+i W^\dag GW}\,.
\end{equation}
$G=(E-H)^{-1}$ is the Green's function of the closed system and matrix $W=(W_a,W_c)$ contains the information on the coupling. As before, index ``$c$''  refers here to the antenna with variable coupling, and ``$a$'' to the measuring antenna. Per definition, the $S$-matrix relates the amplitudes of the incoming ($u$) and outgoing ($v$) waves,
\begin{equation}\label{eq:s03}
    S\left(\begin{array}{c}
              u_c \\
              u_a
            \end{array}\right)
    =\left(\begin{array}{c}
              v_c \\
              v_a
            \end{array}\right).
\end{equation}
A termination of antenna $c$ is described by
\begin{equation}\label{eq:s04}
    u_c=rv_c\,,\qquad r=e^{-(\alpha-i\varphi)}\,,
\end{equation}
where $r$ contains the information on the reflection properties of the antenna. For reflection at an antenna with open or closed end we have $\alpha=0$ (as long as the absorption in the antenna can be neglected). The termination of the antenna by  a 50\,$\Omega$ load corresponds to $\alpha\to\infty$.

Making use of Eq.~(\ref{eq:s02}), one can rewrite Eq.~(\ref{eq:s03}) as
\begin{equation}\label{eq:s05}
i W^\dag GW\left(\begin{array}{c}
              u_c+v_c \\
              u_a+v_a
            \end{array}\right)
    =\left(\begin{array}{c}
              u_c-v_c \\
              u_a-v_a
            \end{array}\right).
\end{equation}
Substituting relation (\ref{eq:s04}) in Eq.~(\ref{eq:s05}), $u_c$ and $v_c$ can be eliminated, resulting in an equation for $u_a$ and $v_a$,
\begin{equation}\label{eq:s06}
    iW_a^\dag\hat{G}W_a(u_a+v_a)=u_a-v_a.
\end{equation}
Here, we have introduced the modified Green's function, $\hat{G}$, with the following matrix element
\begin{equation}\label{eq:s09}
    W_a^\dag\hat{G}W_a \equiv G_{aa}-G_{ac}\frac{1}{1+i\lambda_T G_{cc}}i\lambda_T G_{ca}\,,
\end{equation}
where $G_{nm}=W^\dag_n G W_m$ and $\lambda_T$ is the coupling constant of the ``terminator,''
\begin{equation}\label{eq:s08}
    \lambda_T=\frac{1-r}{1+r}=\tanh\frac{\alpha+i\phi}{2}\,.
\end{equation}

Equation~(\ref{eq:s06}) has the same form as Eq.~(\ref{eq:s05}), but for the measuring antenna only and with the modified Green's function. Substituting explicit expressions for matrix elements $G_{nm}$, we obtain in a number of elementary steps
\begin{equation}\label{eq:s10}
  \hat{G} = G \frac{1}{1+i\lambda_T W_c W_c^\dag G} \equiv \frac{1}{E-H^a_\mathrm{eff}}\,,
\end{equation}
where  $H^a_\mathrm{eff}=H-i\lambda_T W_cW_c^\dag$. Introducing the normalized coupling vector $V=\frac{1}{\sqrt{\lambda_W}}W_c$, where $\lambda_W=W_c^\dag W_c$ is a channel coupling strength, $H^a_\mathrm{eff}$ may be finally written as
\begin{equation}\label{eq:s12a}
    H^a_{\rm eff}=H-i\lambda VV^\dag, \quad \lambda=\lambda_T\lambda_W\,.
\end{equation}
The total coupling constant $\lambda$ is generally complex and takes into account the effects of both the channel coupling ($\lambda_W$) and the terminator ($\lambda_T$). The $2\times2$ scattering matrix [Eq.~(\ref{eq:s02})] for the measuring antenna and the antenna with variable terminator has thus been reduced to a $1\times1$ scattering matrix for the measuring antenna only,
\begin{equation}\label{eq:s12b}
    S_{aa}=\frac{1-i W_a^\dag\displaystyle\frac{1}{E-H^a_\mathrm{eff}} W_a}{1+i W_a^\dag \displaystyle\frac{1}{E-H^a_\mathrm{eff}}W_a}.
\end{equation}
In the case of a single measurement antenna and one antenna with variable coupling, Eq.~(\ref{eq:s12b}) is equivalent to Eq.~(\ref{eq:s_cc2}). Equations~(\ref{eq:s10})--(\ref{eq:s12b}) constitute the main result of this section. They show that the influence of the variable antenna can be taken into account by an appropriate modification of the Hamiltonian.

Two special cases are of particular importance. For the termination of the antenna with a 50\,$\Omega$ load the outgoing wave is completely absorbed, corresponding to the limit $\alpha\to\infty$. It follows $\lambda_T=\tanh \infty=1$ and
\begin{equation}\label{eq:s13}
    H^a_{\rm eff}=H-i\lambda_W VV^\dag\,.
\end{equation}
In this case the coupling is purely imaginary. For the two cases, where the antenna is terminated by a hard wall or an open reflecting end, we may assume $\alpha=0$, resulting in $\lambda_T=\tanh(i\varphi/2)=i\tan\varphi/2$, and
\begin{equation}\label{eq:s14}
    H^a_{\rm eff}=H+\tan\left(\frac{\varphi}{2}\right)\lambda_W VV^\dag\,.
\end{equation}
In this case the coupling is purely real, and the antenna does not correspond any longer to an open channel but to a scattering center only. This is true, as long as the absorption in the antenna can really be neglected. This becomes questionable, as soon as $\varphi$ approaches $\pi$, corresponding to the excitation of a resonance within the antenna. For this singular situation the perturbative treatment of the antenna coupling applied in the derivation looses its justification.

The value of $\varphi$ depends on the length of the antenna in units of the wave length and thus on frequency. But independently of frequency the difference of the phase shift $\varphi$ for the reflection at the open end (oe) and the hard wall (hw), respectively, is always $\pi$. A phase difference of $\pi$ means a replacement of the tangent by the cotangent in Eq.~(\ref{eq:s14}), i.\,e.~the coupling constants $\lambda_T$ for the two situations are related via
\begin{equation}
    \lambda_{T,\rm hw}=1/\lambda_{T,\rm oe}
\end{equation}
With the above introduced total coupling constant $\lambda=\lambda_T\lambda_W$, this may be alternatively be written as
\begin{equation}\label{eq:s15}
    \lambda_{\rm hw}\lambda_{\rm oe}=\lambda_W^2=\lambda_{50\Omega}^2
\end{equation}
since $\lambda_W$ is the coupling constant for the 50\,$\Omega$ load, see Eq.~(\ref{eq:s13}). $\lambda_{\rm hw}$ and $\lambda_{\rm oe}$ denote the total coupling constants for the hard-wall and the open-end reflections. These relations allow for explicit tests of the theory.

\section{Results and Discussion}
\label{sec:results}

In this section we want to discuss the experimental and theoretical results for the coupling fidelity decay under the perturbations (a)--(c) described in Sec.~\ref{sec:exper}. For all results below the system without the varied antenna, corresponding to $\lambda=0$, is chosen as the reference, whereas for the perturbed system the coupling constant is $\lambda'=\lambda_{50\Omega}$, $\lambda_{\rm oe}$, or $\lambda_{\rm hw}$, depending on the terminator.

\begin{figure}[t]
  \includegraphics[width=.95\columnwidth]{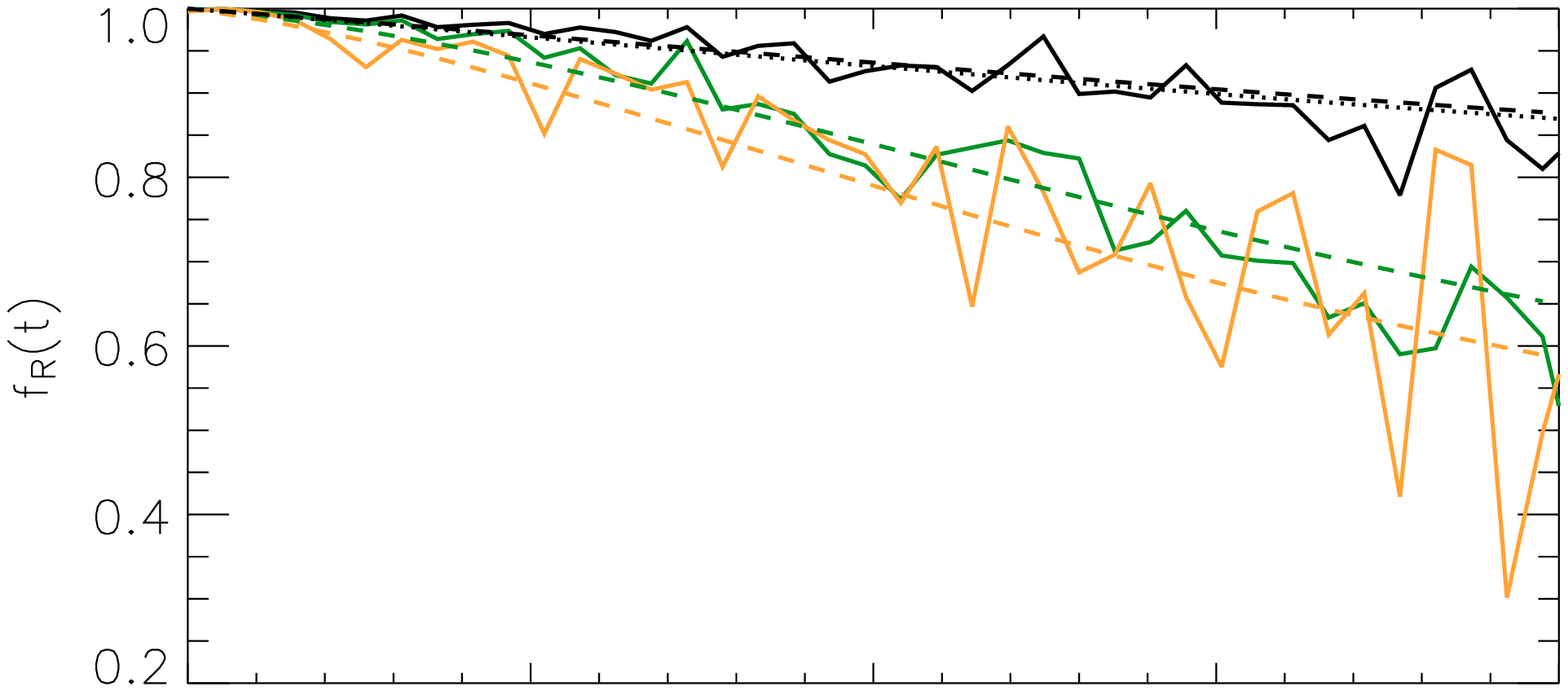}\\[-8ex]
  \includegraphics[width=.95\columnwidth]{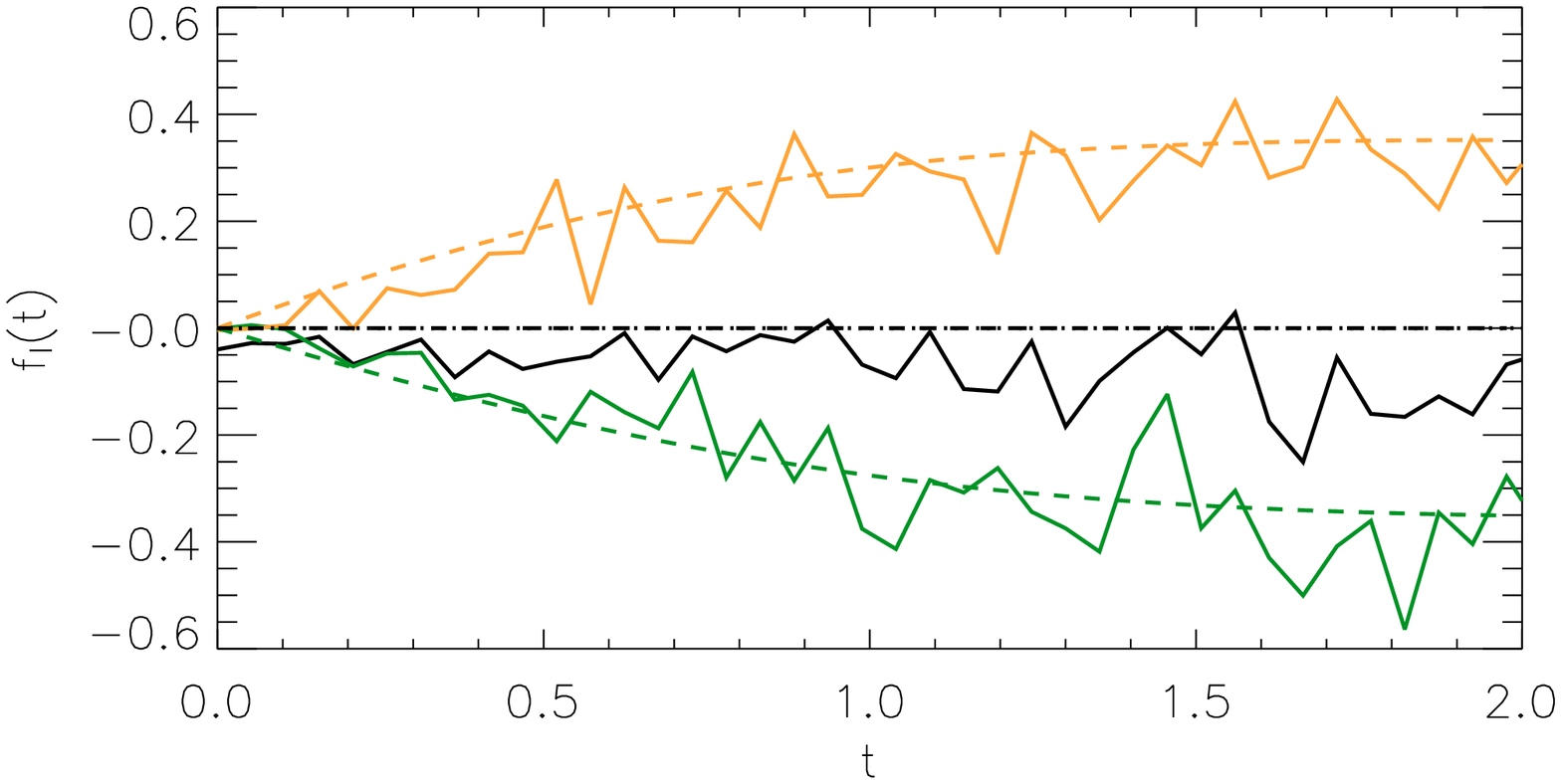}
  \caption{\label{fig:02}(color online)
  Real part $f_R(t)$ and imaginary part $f_I(t)$ of the fidelity amplitude for three types of perturbation: $\lambda_{\rm 50\Omega}$ (black); $\lambda_{\rm hw}$ (orange, light gray); $\lambda_{\rm oe}$ (green, dark gray) in frequency range $8.0-8.5\rm\,GHz$. The time is given in units of the Heisenberg time $t_H=2\pi\hbar/\Delta$, where $\Delta$ is the mean level spacing. Solid lines show the experimental results. The theoretical curves are dotted for experimental parameter (available only for the 50\,$\Omega$ load case) and dashed for fitting parameter. The corresponding parameter and the transmission coefficient for the measuring antenna $a$ are listed in Tab. \ref{tab:01}. The black dotted curve is nearly indistinguishable from the dashed one.}
\end{figure}

We start with a plot of the complex valued fidelity amplitude for one frequency range, see Fig.~\ref{fig:02}. The solid lines show the experimental results, derived from the Fourier transform of the measured $S_{aa}(\nu)$ and $S^{\prime}_{aa}(\nu)$ via relation (\ref{eq:f_ab}) for the situations (a) 50\,$\Omega$ load (black), (b) open-end reflection (green, dark gray) and (c) hard-wall reflection (orange, light gray). For the case (a) we are able to calculate the corresponding theoretical curve (black dotted line) without any free parameter, since the coupling constant $\lambda_W$ can be determined directly from the additional reflection measurement at antenna $c$ (see Sec.~\ref{sec:exper}) using relation (\ref{eq:Tc}). According to Eq.~(\ref{eq:s13}) we expect a purely imaginary coupling with $\lambda_{\rm 50\Omega}$.  For our antenna $\lambda_W$ varies from 0.1 to 0.4 in a range from 6 to $10\,\rm GHz$. Using the experimentally determined $\lambda_W$, one gets already very good agreement between experiment and theory for the $\rm 50\,\Omega$ load without any fit. A fit of $\lambda_W$ to the experimental curves only marginally improves the correspondence.

\begin{figure}
  \includegraphics[width=.95\columnwidth]{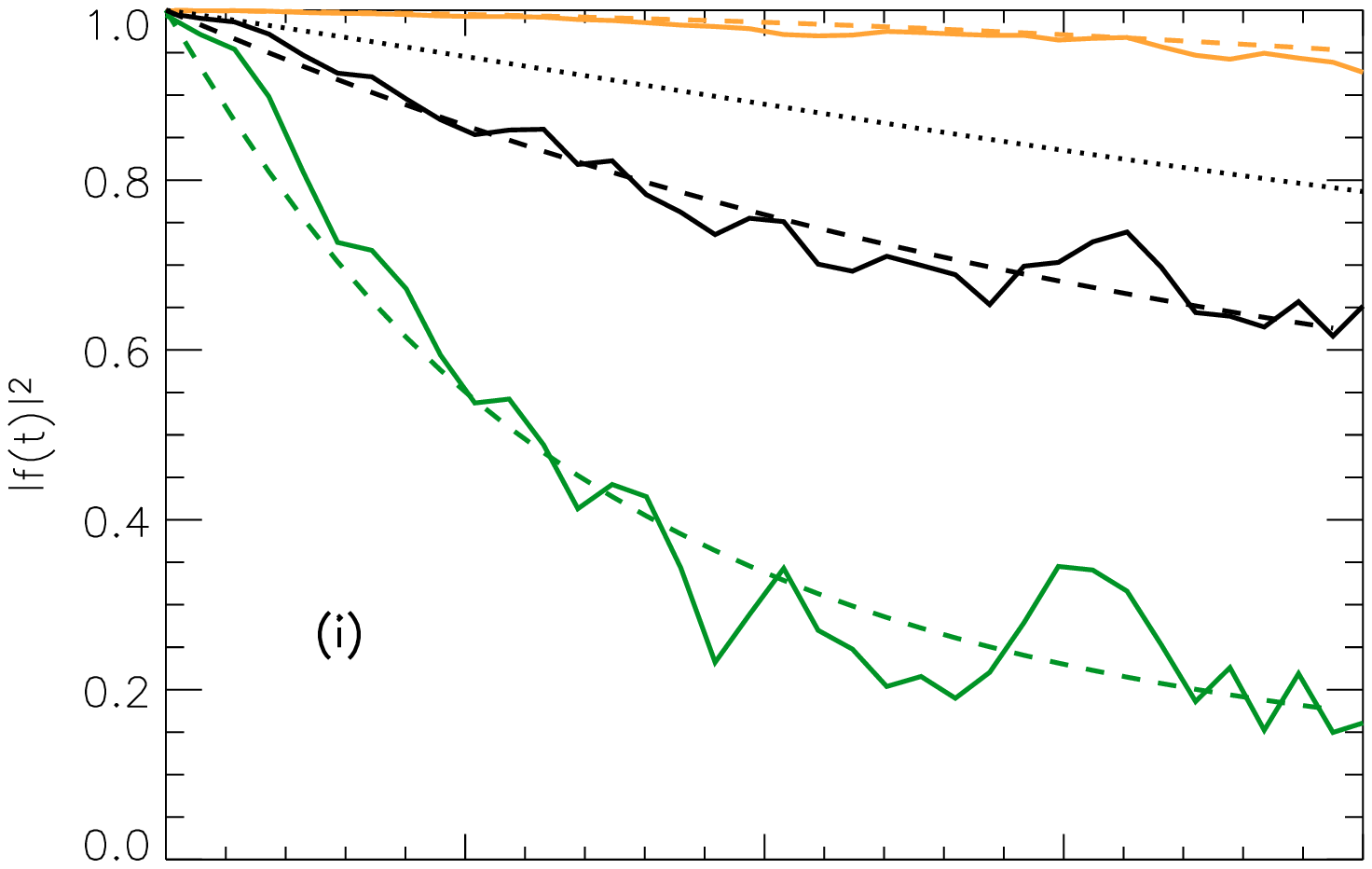}\\[-8ex]
  \includegraphics[width=.95\columnwidth]{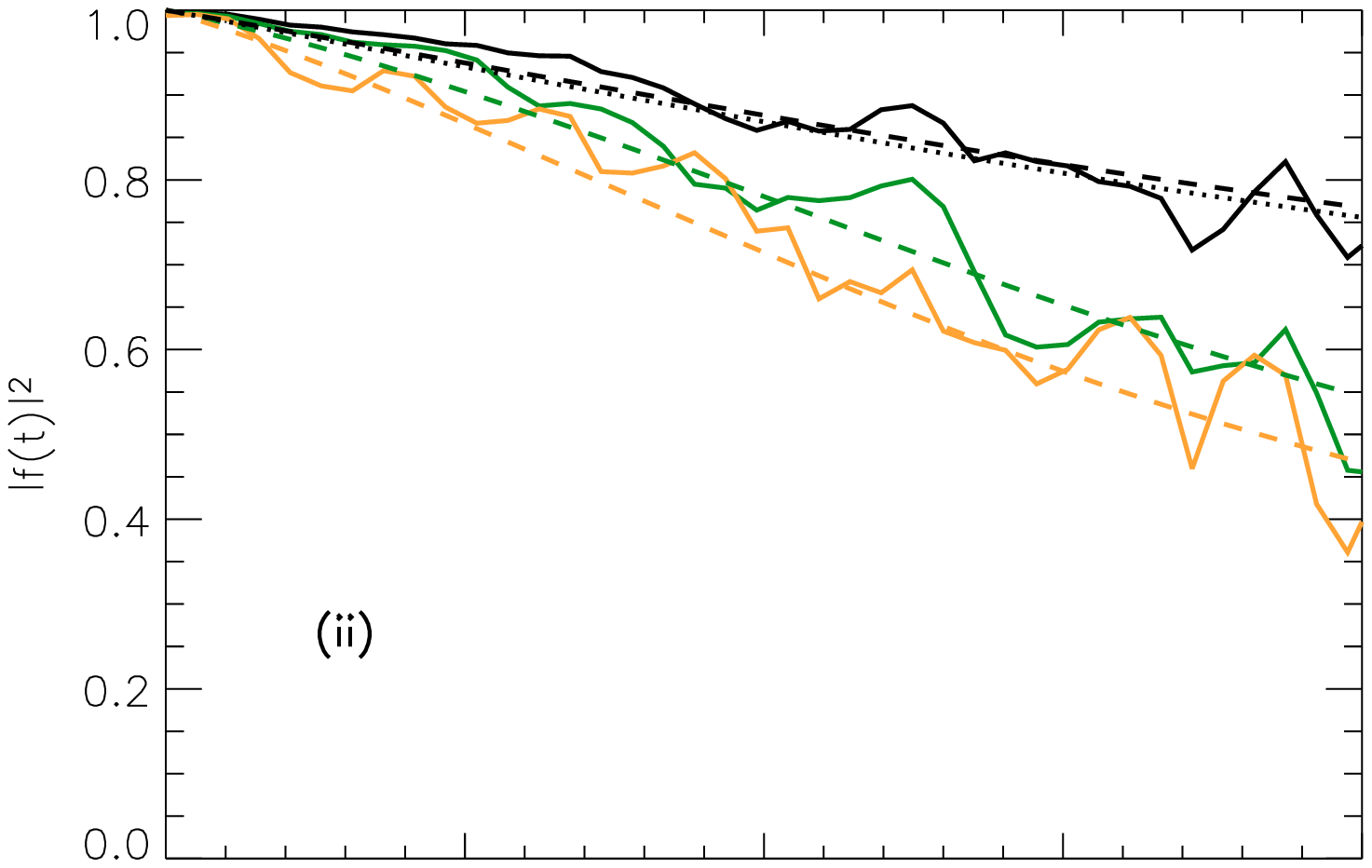}\\[-8ex]
  \includegraphics[width=.95\columnwidth]{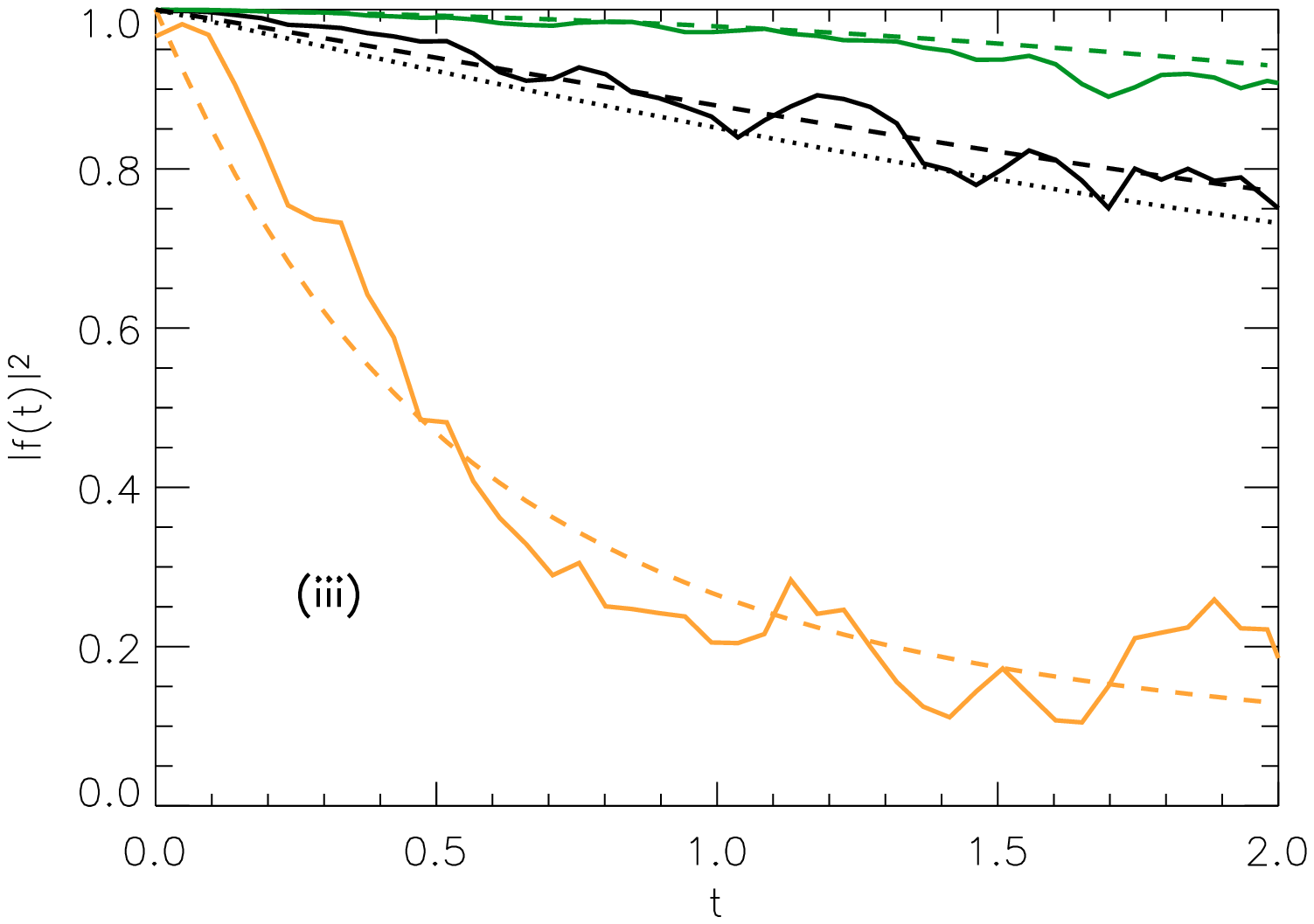}
  \caption{\label{fig:03}(color online)
  Fidelity decay $|f(t)|^2$ for three types of perturbation: $\lambda_{\rm 50\Omega}$ (black); $\lambda_{\rm hw}$ (orange, light gray); $\lambda_{\rm oe}$ (green, dark gray) in three different frequency ranges: (i) $7.2-7.7\rm\,GHz$; (ii) $8.0-8.5\rm\,GHz$; (iii) $8.7-9.2\rm\,GHz$. Solid lines show the experimental results and the theoretical curves are dotted for experimental parameter (available only for the 50\,$\Omega$ load case) and dashed for fitting parameter. The corresponding parameter and the transmission coefficient for the measuring antenna $a$ are listed in Tab.~\ref{tab:01}.}
\end{figure}

For the perturbations with reflecting ends the systems are closed. The correct description for this situations is Eq.~(\ref{eq:s14}). With the total coupling constants $\lambda_{\rm oe}$ and $\lambda_{\rm hw}$ as a free fitting parameter we find again an agreement between experiment and theory (dashed lines), for both, real and imaginary part. As one would expect from theory for the case of reflecting ends we see a significant imaginary part of the fidelity amplitude $f_I(t)$ (green and orange lines), whereas in the case of an absorbing end (black line) the imaginary part is nearly zero.

It is convenient to continue our discussion in terms of the fidelity (not its amplitude) introduced in Eq.~(\ref{eq:F_ab}). In Fig.~\ref{fig:03} we present the experimental and theoretical fidelity results for the situations 50\,$\Omega$ load (black lines), open-end (green lines), and hard-wall (orange lines) as perturbation, for three frequency ranges. In case of the closed channels the total phase shift $\varphi(\nu)$ increases monotonically with frequency. Thus $\lambda_{\rm oe}$ and $\lambda_{\rm hw}$ are oscillating in counter phase, see Eq.~(\ref{eq:s14}). This induces a corresponding oscillation in the strength of the fidelity decay as seen in Fig.~\ref{fig:03}. For a more quantitative discussion we compare the experimentally determined fidelity decay (solid lines) to the theoretical curves (dotted and dashed lines). First of all, one sees that fitting (dashed lines) the experimental results again works well for all cases. Focusing on the 50\,$\Omega$ load case (black lines) also the theoretical results without free parameter (dotted lines) show good agreement with the experiment for the frequency ranges plotted in Figs.~\ref{fig:03}~(ii) and (iii). Only the plot in Fig.~\ref{fig:03}~(i) shows significant deviation between the theoretical result without free parameter and the experimental curve. We want to stress that the case shown here is the worst among all the investigated frequency ranges.
Here the experimental fidelity amplitude $f_{50\Omega}$ shows a significant imaginary part. Thus the imaginary part of $\lambda_{50\Omega}$ is not zero, i.\,e.\ the 50\,$\Omega$ terminator does not correspond to perfect absorption, and Eq.~(\ref{eq:s14}) does not hold. This might be due to an antenna resonance, leading to an increased reflection from the channel $c$.

\begin{table}
  \centering
  \begin{tabular}{c|c|c|c|c|c|c|c}
    &$\nu$/GHz&$\lambda^{\rm exp}_{\rm 50\Omega}$&$\lambda^{\rm fit}_{\rm 50\Omega}$&$\lambda_{\rm oe}$&$\lambda_{\rm hw}$& $\lambda_W$&$T_a$\\
    \hline
    (i)&$7.2-7.7$&$0.19$&$0.37$&$0.65\,\imath$&$-0.04\,\imath$&$0.16$&$0.19$\\
    \hline
    (ii)&$8.0-8.5$&$0.21$&$0.20$&$0.19\,\imath$&$-0.23\,\imath$&$0.21$&$0.22$\\
    \hline
    (iii)&$8.7-9.2$&$0.24$&$0.21$&$0.05\,\imath$&$-0.83\,\imath$&$0.20$&$0.34$\\
  \end{tabular}
  \caption{\label{tab:01}
  Coupling constants in the three different frequency ranges (i)-(iii). According to Eq.~(\ref{eq:s15}), $\lambda^{\rm exp}_{\rm 50\Omega}$ and $\lambda^{\rm fit}_{\rm 50\Omega}$ should be compared to $\lambda_W=\sqrt{\lambda_{\rm oe}\lambda_{\rm hw}}$, see main text for discussion.}
\end{table}

Finally we perform a check on the coupling constants based on Eq.~(\ref{eq:s15}). Accordingly, the square root of the product of the coupling constants for open-end and hard-wall reflection should give $\lambda_{50\Omega}$. Table~\ref{tab:01} shows that for the frequency ranges (ii) and (iii) there is indeed good agreement between $\sqrt{\lambda_{\rm oe}\lambda_{\rm hw}}$, $\lambda^{\rm fit}_{\rm 50\Omega}$ and $\lambda^{\rm exp}_{\rm 50\Omega}$. In the case (i) $\sqrt{\lambda_{\rm oe}\lambda_{\rm hw}}$ agrees quite good with the experimental parameter $\lambda^{\rm exp}_{\rm 50\Omega}$, but the fitting parameter is much larger. This deviation reconfirms our arguments presented in the above discussion of the fidelity plot shown on Fig.~\ref{fig:03}(i).

\section{Conclusions}
\label{sec:conclusions}

In this work, we have studied the influence of the coupling to the continuum on the decay of fidelity. This complements previous experiments of our group, where the fidelity decay under the influence of various types of geometrical perturbations was studied \cite{sch05b,hoeh08a,sch05d,bod09a} but for closed systems exclusively. To get rid of an overall absorption we used the concept of scattering fidelity introduced by us previously \cite{sch05b}, defined as the parametric cross-correlation function of $S$-matrix elements normalized to the corresponding autocorrelation function.

On the theoretical side we have developed a model description of the fidelity decay in terms of a modified VWZ approach. The parametric cross-correlation function of $S$-matrix elements for two different $\lambda \ne \lambda'$ can be reduced to an autocorrelation function with a complex effective transmission coefficient [Eq.~(\ref{eq:Teff})], thus expressing coupling fidelity in terms of a modified VWZ integral. This theory holds for an arbitrary number of channels and describes the experimental coupling fidelity results well. It would be interesting to investigate, whether it is possible to relate coupling fidelity obtained via the VWZ ansatz to an approximation using the random plane wave conjecture, as used to explain the local fidelity \cite{hoeh08a}.

We have found two additional important result. First, a smooth variation of the coupling, e.\,g.\ by varying the coupling to an attached wave guide will not easily yield the information about the effect of coupling to the continuum on the scattering fidelity. Each geometric variation will give rise to both a change of coupling and internal scattering properties, thus screening the purely external effect, as is also discussed in Appendix~\ref{app:Exp}.

Second, we have included closed channels within the description of VWZ. The speed of the fidelity decay for the open-end and hard-wall reflection oscillates with frequency due to the corresponding variation of the phase with frequency. An important relation (\ref{eq:s15}) between the coupling constants of the antenna terminations has been established, enabling us to connect the results found for the closed channel ($\lambda_{\rm oe}$, $\lambda_{\rm hw}$) to those for the open channel ($\lambda_{50\Omega}$). In all cases the fidelity decay for at least one of the reflecting antennas is faster than for the open channel, showing the strong influence of the imaginary part on the coupling constant $\lambda$.

\begin{acknowledgments}
One of us (D.V.S.) is grateful to I.~Smolyarenko for useful discussions of the results. We would like to acknowledge the generous hospitality of CIC (Cuernavaca, Mexico) during our stay there at the program RMT-MEX09, where this work has been partly completed. The experiments have been founded by the Deutsche Forschungsgemeinschaft via the research group 760 ``Scattering systems with complex dynamics''. T.\,H.\,S. thanks the DFG for support of a number of visits in Marburg. T.\,H.\,S. and T.\,G.\ have been supported by CONACyT under Grant No.~79988 and by PAPIIT, Universidad Nacional Aut\'{o}noma de M\'{e}xico under Grant No.~IN-111607 and IN-114310.
\end{acknowledgments}

\appendix

\section{Experiment with attached wave guide with variable coupling}
\label{app:Exp}

\begin{figure}[t]
  \includegraphics[width=.95\columnwidth]{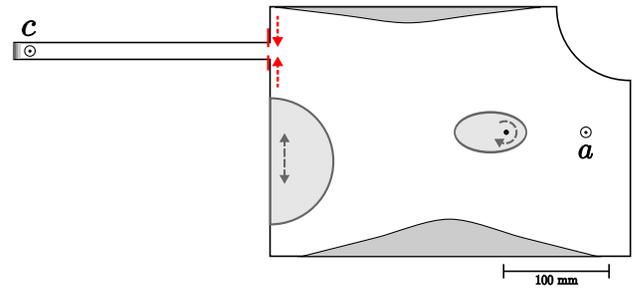}
  \caption{\label{fig:04}
  Geometry of the billiard with attached wave guide. }
\end{figure}

In our first approach we used the setup shown in Fig.~\ref{fig:04} where the opening of the variable slit plays the role of the fidelity parameter. The setup is based on a quarter Sinai shaped billiard with length $l=\rm 342\,mm$, width $w=\rm 237\,mm$ and a quarter-circle of radius $r =\rm 70\, mm$, and an attached channel. The channel has a total length $l_c=\rm 243\,mm$ and a width $w_c=\rm16\,mm$. At position $a$ and $c$ two antennas were fixed and connected to the VNA. The complete $S$-matrix was measured in a frequency range from $\rm 9.5$ to $\rm 18.0\,GHz$ with a resolution of $\rm 0.1\,MHz$, where the wave guide only supports a single propagating mode, i.\,e.~it acts as a single channel. The perturbation of the system was achieved by opening the channel from $d=0-16$\,mm in steps of 0.1\,mm using a slit diaphragm at the point of attachment. An ellipse insert with semiaxis $a=\rm 70\,mm$ and $b=40$\,mm was rotated to get an ensemble of 20 different systems for averaging. Additional elements were inserted into the billiard to avoid bouncing-ball resonances. The wave guide was terminated by a perfect absorber, which according to Eq.~(\ref{eq:s13}) should correspond to a purely imaginary coupling. As before the coupling constant $\lambda_W$ could be determined directly from a reflection measurement at antenna $c$.
$\lambda_W$ could be varied from $\lambda_W=0$ (no coupling) to $\lambda_W=1$ (perfect coupling) by increasing the opening $d$ of the slit.

\begin{figure}[b]
  \includegraphics[width=.98\columnwidth]{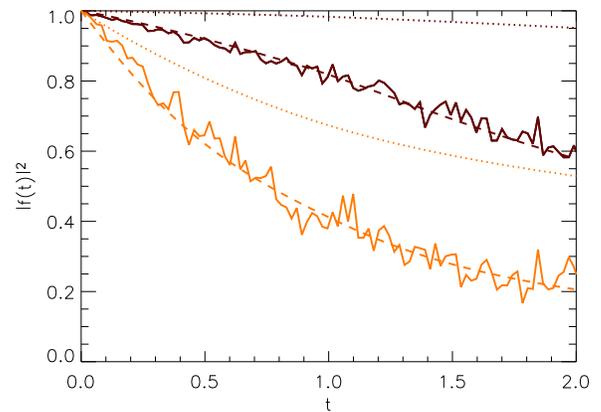}
  \caption{\label{fig:05}(color online)
  Experimental coupling fidelity $|f(t)|^2$ (solid lines) and theoretical results for the experimental parameter $\lambda_W$ (dotted line) and the fit parameter $\lambda_{\rm fit}$ (dashed line) for two openings $d=\rm 6.5\,mm$ with $\lambda_W=0.05$ and $\lambda_{\rm fit}=-0.18i$ (black), and $d=\rm 11.2\,mm$ with $\lambda_W=0.52$ and $\lambda_{\rm fit}=-0.55i$ (orange, light gray). The frequency window of the Fourier transform of the measured $S_{\rm aa}(\nu)$ was 13 to $\rm14\,GHz$ and the transmission coefficient for antenna $a$ was $T_a=0.95$.}
\end{figure}
In Fig.~\ref{fig:05} the coupling fidelity decay is shown for two different perturbation strengths. The solid lines show the experimental results. With the formulas derived in Sec.~\ref{sec:theory}, we  calculated the expected theoretical fidelity decay assuming that the channel is totally open, i.\,e.~the coupling is purely imaginary (dotted lines). There is obviously no agreement. This shows that something is wrong in the argumentation. For a further check we removed the absorbing end and the antenna in the channel and replaced it by a reflecting end thus closing the system. We did not find any noticeable difference to the case with the absorbing end and the antenna in the channel experimentally. So there is only one explanation: by far the major part of the wave is reflected directly at the slit, and only a minor part really penetrates into the channel.  This means that the coupling is not imaginary but mainly real (up to perhaps a minor imaginary contribution), and we should use Eq.~(\ref{eq:s14}) instead of Eq.~(\ref{eq:s13}) for the interpretation of our results. The dashed lines in Fig.~\ref{fig:05} show the resulting theoretical curves with $\lambda_{\rm fit}$ as a free parameter according to the definition preceding Eq.~(\ref{eq:s14}). Now a perfect agreement between experiment and theory is found.

As a resume we can state that the variable slit works essentially as a scattering center leading to partial masking of the change of coupling by the change of scattering properties in the fidelity decay.

\section{Derivation of EQ.~(\ref{eq:Teff}) and qualitative discussion}
\label{app:theo}

The calculation of Eq.~(\ref{eq:ss_ft}) proceeds along the same lines as in \cite{ver85a}, hence we indicate below only the main steps and essential differences. First, we make use of the representation of resolvents and thus $S$-matrix elements [Eq.~(\ref{eq:s_cc1})] in terms of Gaussian integrals over auxiliary ``supervectors'' consisting of both commuting and anticommuting (Grassmann) variables. This allows us to perform statistical averaging over GOE exactly. Then in the RMT limit $N\to\infty$, the remaining integration over the auxiliary field can be done in the saddle-point approximation. It turns out that there exists a nontrivial saddle-point manifold \cite{efe83} over which one has to integrate exactly. As a result, the two-point correlation function  of the $S$-matrix elements in the energy domain acquires the form of a certain expectation value in field theory (nonlinear supersymmetric $\sigma$-model), $\langle(\cdots)\rangle=\int\mathrm{d}[\sigma_G]e^{\mathcal{L}(\varepsilon)}\mathcal{F}_{M}(\cdots)$,  cf. Eq.~(7.13) of Ref. \cite{ver85a}. In the notations of this paper, the effective Lagrangian reads $\mathcal{L}(\varepsilon)=\frac{1}{4}N\varepsilon\,\mathrm{trg}(\sigma_GL)$, with $\varepsilon$ being the energy difference. Definitions of the supertrace, $\mathrm{trg}$, as well as of the supermatrices $\sigma_G$ and $L$ can also be found there (see \cite{efe96} for a general reference). The pre-exponential terms omitted above depend on the coupling constants in the channels $a$ and $b$ ($a,b\neq c$), being thus the same as considered in \cite{ver85a}. They finally  correspond to the expressions appearing explicitly in Eqs.~(\ref{eq:J}) and (\ref{eq:P}). At last, the so-called channel factor $\mathcal{F}_M$ accounts for the coupling to all the channels. It is the term that requires modification due to both generally complex and varied coupling constants. In the [1,2] block notation (the ``advanced-retarded'' ordering of supermatrix elements), $\mathcal{F}_M$ reads
$$
 \mathcal{F}_M = \prod_{k=1}^M \exp\left\{-\frac{1}{2} \mathrm{trg} \ln \left[ 1 + i
   \left(\begin{array}{cc} \lambda_k & 0 \\ 0 & \lambda_k'^{*} \end{array}\right)
 \sigma_G L \right]\right\},
$$
where $\lambda_k=\lambda_k'$ for all channels save the varied one, $k\neq c$.  By employing the ``angular'' parametrization of $\sigma_{G}$ in terms of the matrices $t_{12}$ and $t_{21}$, the subsequent evaluation of $\mathcal{F}_M$ goes in parallel with Sec. 7 of Ref.~\cite{ver85a}, with the final result being
\begin{equation}\label{eq:app1}
 \mathcal{F}_M = \prod_{k=1}^M \exp \left[-\frac{1}{2} \mathrm{trg} \ln \left(
  1 + T_k^{\mathrm{eff}} t_{12}t_{21} \right)\right]\,.
\end{equation}
This is just a usual formula for the channel factor in the VWZ theory except for the effective transmission coefficient $T_c^{\mathrm{eff}}$ in the channel $c$ that is now given by expression (\ref{eq:Teff}) (we note that $T_k^{\mathrm{eff}}=T_k$ if $k\neq c$) \cite{note2}. Performing finally the Fourier transform, the two-point correlation function in the time domain takes the form of Eqs.~(\ref{eq:VWZ})--(\ref{eq:I}), with the above modification in the channel factor corresponding explicitly to the second line of Eq.~(\ref{eq:I}).

Although the subsequent evaluation cannot be made analytically and has to be done numerically, it is still useful to make some qualitative analysis. To this end we note that $P_{ab}(t)$ and $J_a(t)$ are quite similar in structure to the ``norm leakage'' decay function \cite{sav97} and the form factor of the Wigner time delays \cite{leh95b}. Following the analysis performed there (see also \cite{dit00}), one notices that the time dependence in question is mainly due to the channel factor [Eq.~(\ref{eq:app1})].  In the time domain, its typical behavior is  $\sim\prod_{k=1}^M(1+\frac{2}{\beta}T_k t)^{-\beta/2}$, where $\beta=1$ is for the present case of time-invariant systems whereas $\beta=2$ is for the case of broken time invariance (GUE). The case of the coupling variation in the channel $c$ amounts then to replacing in this expression $T_c$ with $T_c^{\mathrm{eff}}$. This suggests the following heuristic formula for the coupling fidelity
\begin{equation}\label{eq:app2}
 F_{\mathrm{surm}}(t) = \left[\frac{(1+2T_c t/\beta)(1+2T'_c t/\beta) }{
                         |1+2T^{\rm eff}_c t/\beta|^2}\right]^{\beta/2}\,.
\end{equation}
For the parameters of $\lambda$ and $\lambda^\prime$ found in the experiment only deviations on the \% level were found while making fit to surmise (\ref{eq:app2}). We stress, however, that there is no control on approximations involved to derive this expression. One should generally expect that surmise (\ref{eq:app2}) coincides closely with the exact result at small times (when it is given by an exponential dependence), while the exact asymptotic behavior at large times is reproduced up to a factor of the order of unity (as was indeed confirmed numerically). Therefore, we have used the exact supersymmetry result for all the figures and analysis of the main text.


\begin{thebibliography}{42}
\bibitem{note1}
  \bibinfo{note} We omit the overall factor equal to the length of the energy
  spectrum used to compute the Fourier transforms in Eq.~(\ref{eq:ss_ft}), since
  it is finally canceled due to the normalization in definition (\ref{eq:f_ab}).
\bibitem{note2}
  \bibinfo{note} It is worth noting that in the case of the pure imaginary
  $\lambda $, i.\,e.\ a closed channel, expression (\ref{eq:app1}) is
  reduced to the action derived in \cite{mar03} for the case of localized
  perturbations in closed systems, $T_c^\mathrm {eff}$ being then
  expressed in terms of the reactance matrix of the scattering centre.
\end{thebibliography}
\end{document}